\documentclass[fleqn,usenatbib]{mnras}

\usepackage{newtxtext,newtxmath}

\usepackage[T1]{fontenc}

\DeclareRobustCommand{\VAN}[3]{#2}
\let\VANthebibliography\thebibliography
\def\thebibliography{\DeclareRobustCommand{\VAN}[3]{##3}\VANthebibliography}

\usepackage{graphicx}	
\usepackage{amsmath}	
\usepackage{gensymb}

\title[Diffuse X-ray Background Measured with HXMT] {Insight-HXMT Measurements of the Diffuse X-ray Background}
\author[Rui, Huang et al.]{%
    R. Huang$^{1}$ \thanks{E-mail: hr17@mails.tsinghua.edu.cn},
    W. Cui$^{1}$ \thanks{E-mail: cui@tsinghua.edu.cn},
    J. Y. Liao$^{2}$,
    S. Zhang$^{2}$,
    S. F. Wang$^{1}$,
    J. Jin$^{2}$,
    X. F. Lu$^{2}$,
    C. C. Guo$^{2,3}$,
    \newauthor
    Y. You$^{2}$,
    G. Li$^{2}$,
    J. Zhang$^{2}$,
    \\
\\
\\
$^{1}$ Department of Astronomy, Tsinghua University, Beijing 100084, China\\
$^{2}$ Key Laboratory for Particle Astrophysics, Institute of High Energy Physics, Beijing 100049, China\\
$^{3}$ University of Chinese Academy of Sciences, Chinese Academy of Sciences, Beijing 100049, People’s Republic of China\\} 

\date{Accepted 20 Apri 2022}

\pubyear{2021}

\begin{document}
\label{firstpage}
\pagerange{\pageref{firstpage}--\pageref{lastpage}}
\maketitle

\begin{abstract}
We present an X-ray spectrum of the diffuse X-ray background (DXRB) between 1.5 and 120 keV, as measured with the Low-Energy Detector (LE) and the High-Energy Detector (HE) aboard the Insight-HXMT satellite, based on 'blank-sky' observations. LE covers a nominal energy range of 1-15 keV and HE 20-250 keV, but calibration issues and data quality narrowed the energy range for this work. The LE background was directly measured with `blind' detector modules, while the HE background was derived from Earth-occultation data. 
With the LE data alone, the measured DXRB spectrum can be well described by a power law; fitting the LE and HE data jointly, however, a spectral cut-off must be introduced in the model to account for the measurements above 30 keV. Modelling the combined spectrum with a cut-off power law, the best-fit photon index is 1.40, normalisation $9.57$~$\rm ph~cm^{-2}~s^{-1}~keV^{-1}~sr^{-1} $ (at 1 keV), and cut-off energy 55 keV, after correcting for the effects of the Earth albedo and atmospheric emission (which are significant in the HE band). Based on the best-fit cut-off power law, we derived the spectral energy distribution (SED) of the DXRB. The shape of the SED is in general agreement with the published measurements, but the overall normalization is lower by varying amounts, except for the HEAO-1 result, with which our result is in good agreement.


\end{abstract}


\section{Introduction}
\label{Intro}
The presence of diffuse X-ray background (DXRB) radiation is among the first discoveries in X-ray astronomy (\citealt{giacconiEVIDENCERAYSSOURCES1962}). 
Over the past decades, the spectrum of the DXRB has been well characterized at low energies (e.g., \citealt{lumbXrayBackgroundMeasurements2002},\citealt{revnivtsevSpectrumCosmicXray2003} and \citealt{cappellutiChandraCOSMOSLegacy2017}), 
but much less so at high energies (above about 25 keV), due to limited observing capabilities. It is generally agreed that the X-ray spectrum of the DXRB is of power-law shape at low energies, with a photon index of about 1.4, and that the spectrum rolls over at high energies (\citealt{gruberSpectrumDiffuseCosmic1999b}), although the normalisation of the power law and the roll-over energy still have significant measurement uncertainties. Although not entirely understood, the DXRB is generally thought to originate mostly from the integrated X-ray emission of active galactic nuclei (AGN) at energies above roughly 1 keV (e.g.,\citealt{hasingerXray1993}; \citealt{cappellutiChandraCOSMOSLegacy2017}), but the exact fraction of the AGN contribution depends critically on the normalization of the DXRB spectrum. The latter could have important cosmological implications, as the hot gas that is thought to permeate the large-scale structures in the present-day universe (or in the halo of galaxies) could also contribute to the DXRB (\citealt{croftHydrodynamicSimulationCosmological2001}). Going down in energies, there is consensus on the increasing contribution to the DXRB from the hot gas associated with the Milky Way, possibly in the form of a `Local Bubble' surrounding the Sun and a Galactic halo (\citealt{mccammonSoftXrayBackground1990}; \citealt{snowdenImplicationsROSATObservations1993}), but with a significant component associated with the solar-wind charge exchange processes (\citealt{snowdenXMMNewtonObservationSolar2004},\citealt{galeazziDXLSoundingRocket2011}).

In this work, we present a measurement of the DXRB in the energy range of 1.5-120 keV, with the Low-Energy Detector (LE) and the High-Energy Detector (HE) aboard the Hard X-ray Modulation Telescope (also known as Insight-HXMT) X-ray satellite, as part of the efforts to characterize cosmic and non-cosmic background levels of the detectors.

\section{Observations}
\label{obser}
Insight-HXMT is the first Chinese satellite that is dedicated to X-ray astronomy (\citealt{liHXMTChineseHighenergy2007}; \citealt{zhangIntroductionHardXray2014}; \citealt{zhangOverviewHardXray2020}). It was launched on June 15, 2017, and has been in successful operation ever since. Insight-HXMT carries three mechanically-collimated science instruments aboard: besides LE and HE, there is also the Medium-Energy Detector (ME). 

HE is the primary payload, situated in the middle of the assembly layout (\citealt{zhangInsightHXMTMissionIts2018};\citealt{liuHighEnergyXray2020}). It consists of 18 identical NaI(Tl)/CsI(Na) phoswich detector modules, covering a nominal energy range of 20-250 keV, with a collective geometrical area of about 5000 $\mathrm{cm^2}$. Sixteen of the detector modules are collimated to a field-of-view (FoV) of $1.1 \degree \times 5.7 \degree $ , with one being `blind' (i.e., with a thick cover), and two to an FoV of $5.7 \degree \times 5.7 \degree$. This choice of the FoV combination and `blind' detector was made to allow a more effective characterization of the detector background.

Surrounding HE are ME and LE, which follow a similar design philosophy. ME consists of 54 modules of Si PIN detectors, covering a nominal energy range of 5-30 keV, with a collective geometrical area of about 950 $cm^2$. Forty eight of the detector modules have a FoV of $1\degree \times 4\degree$, with three being `blind', and six have a FoV of $4 \degree \times 4\degree$. LE (\citealt{chenLowEnergyXray2020} ) consists of 90 modules of swept-charge devices, covering a nominal energy range of 1-15 keV, with a collective geometrical area of about 380 $cm^2$. Sixty-three of the detector modules have a FoV of $1.6\degree\times 6\degree$, with three being `blind', twenty-one have a FoV of $6\degree\times 6\degree$, with three being `blind',  and six have a very large FoV of $60\degree \times 2.5\degree$. Both for ME and LE, the detector modules were assembled into three separate units.

Here, we used data from observations of the `blank-sky' fields (free of any known bright X-ray sources, including Proposal IDs P0101293, P0202041, P0301293), which were conducted specifically for quantifying detector background. Because the performance of ME was unstable, only the HE and LE data were used for this study. 

\section{Data reduction}
The Level-1 data were further filtered, processed and analyzed with the standard data analysis package, HXMTDAS (v2.04). In the following, the procedures are described in detail separately for LE and HE data.

\subsection{LE data}
\label{LE}
\begin{figure}
\begin{center}
\includegraphics[width=1.0 \columnwidth]{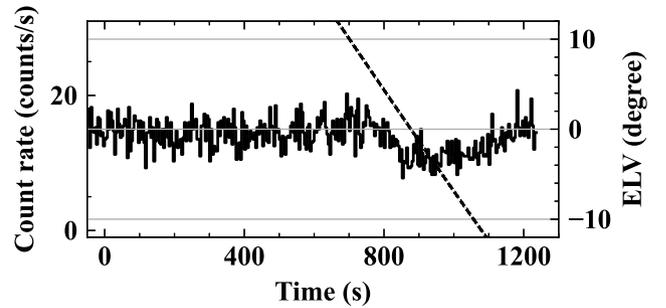}
\end{center}
\caption{The LE light curve of a `blank-sky' field. The data were taken from ObsID P0101293152. Also shown (in dashed black line) is the variation in ELV through an Earth occultation. Note the decrease in the count rate when the field is occulted by the Earth (with ELV turning negative).} 
\label{lc_of_LE}
\end{figure}

For data filtering, the relevant parameters include: elevation angle (of the pointing direction) above the Earth limb ($ELV$), elevation angle above the {\it bright} Earth limb ($DYE\_ELV$), geomagnetic cutoff rigidity ($COR$), offset angle from the pointing direction ($ANG\_DIST$), South Atlantic Anomaly flag ($SAA\_FLAG$), elapsed time after an SAA passage ($TN\_SAA$), time prior to an SAA passage ($T\_SAA$), offset angle from the Sun ($SUN\_ANG$), and offset angle from the Moon ($MOON\_ANG$). Here, we adopted the following criteria (\citealt{guanPhysicalOriginNonphysical2021}) to obtain good time intervals (GTIs): $ELV>10$ , $DYE\_ELV=200$ (indicating that the observation is conducted when the spacecraft is on the dark side of the Earth), $COR>8$, $ANG\_DIST<0.04 $, $SAA\_FLAG=0 $, $T\_SAA>300$, $TN\_SAA>300$, $SUN\_ANG>30$, and $MOON\_ANG>30$. Using the filtered data, we made a light curve and a spectrum for each observation, combining all detector modules of small FoV, as well as the corresponding response matrix file (RMF) and auxiliary response file (ARF) for spectral analyses. 
Fig.~\ref{lc_of_LE} shows a light curve of a `blank-sky' field observed with LE around the time of an Earth occultation. The light curve shows a clear decrease in the count rate at around 800 s, indicating the beginning of blockage of the FoV by the Earth. For spectral analyses, we experimented with two different methods of constructing background spectra: (1) using data from the `Earth-occulted' portion of an observation, and (2) using data from the blind detectors. We found that the former would suffer from systematic effects, which might be related to light leak through filters on LE, and concluded that the latter would be more accurate. We did not simply take the count rate of the blind detectors as background; instead, we found correlations in count rate between the blind detectors and other detectors during the times of earth occultation and used them to derive a background spectrum for each normal detector, assuming the particle background is of the same spectral shape in blind detectors as in normal ones (\citealt{liaoBackgroundModelLowenergy2020}). 

Combining data from small-FoV detectors, we made a spectrum from a `blank-sky' observation and subtracted from it the corresponding background spectrum to obtain a spectrum of the DXRB, To improve the statistical precision of the DXRB measurement, we repeated the procedure for all `blank-sky' observations, which had been conducted before July 2019 (i.e., the first HXMT data release), and stacked the extracted source and background spectra separately. Note that we only used data from early observations in this work, because of the worsening performance of LE later on. In the end, we accumulated an effective exposure time of $117\ ks$. Accordingly, we also co-added the corresponding instrumental response files (RMFs and ARFs) for subsequent spectral modelling.

\subsection{HE data}
\label{HE}
The GTIs for the HE observations were derived based almost on the same criteria as LE: $DYE\_ELV=200$, $COR>8$, $ANG\_DIST<0.04 $, $SAA\_FLAG=0 $, $T\_SAA>300$, $TN\_SAA>300$, $SUN\_ANG>30$, $MOON\_ANG>30$. Unlike in the case of LE, however, we were not able to use the blind detectors to measure the particle background for two reasons. First of all, the energy resolution is significantly different among the HE detectors. This is important because HE spectra show strong instrumental lines that are produced in the collimators (e.g. 31, 57, 65, and 191 keV lines, see \citealt{zhangComparisonSimulatedBackgrounds2020}). The difference in energy resolution would lead to significant residuals around the energies of the lines, if the spectrum made with data from the blind detector is treated as background. Secondly, the extra cover of blind detectors could also introduce discrepancies in the background spectrum, when compared with the normal detectors. In this work, therefore, we used `Earth-occulted' observations to model the detector background for HE.

From the `blank-sky' observations conducted during the first two years of the mission, we collected 166 segments of Earth-occultation transitions, with the $ELV$ of each segment nominally covering a range between -10 and 15. For each transition, we made a light curve for each of the HE detector modules, and examined the transition into or out of an earth occultation. For HE, the DXRB signal is quite weak and thus yields much lower signal-to-noise ratio (SNR) than for LE. To improve statistics, we proceeded to stack data from individual observations.

\begin{figure}
\begin{center}
\includegraphics[width=1.0 \columnwidth]{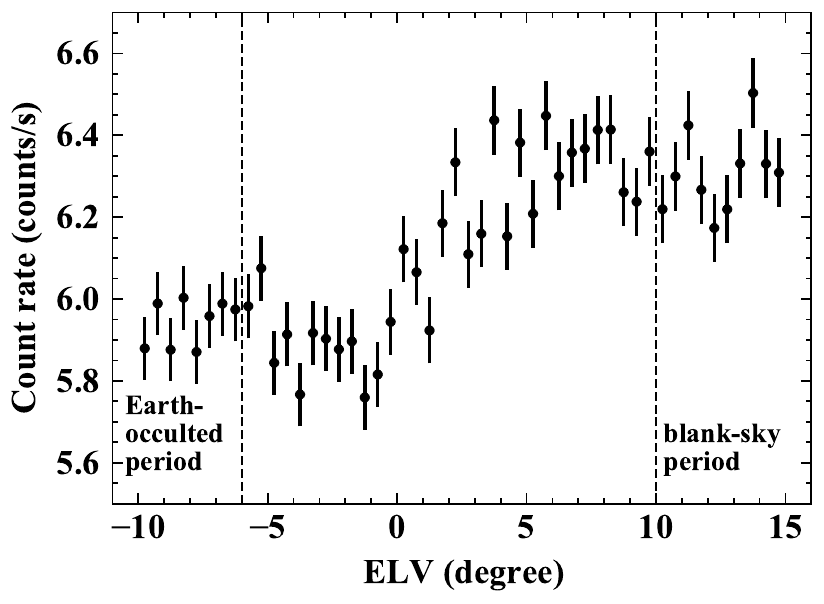}
\end{center}
\caption{Stacked HE count rates as a function of ELV. It is based on data from the two large-FoV detectors, covering energy channels between 26 and 35. Note the increase in count rate as ELV turns positive. The vertical dashed lines indicate the boundaries of the defined "Earth-occulted" and "blank-sky" periods.} 
\label{HElc}
\end{figure}

Fig.~\ref{HElc} shows the stacked light curves based on the `blank-sky' observations with the two large-FoV detector modules. The presence of the DXRB signal is indicated by the increase in count rate outside the Earth occultation. Based on the light curve, we defined the `Earth-occulted' and `blank-sky' time periods as $ELV < -6$ and $ELV>10$, respectively, for subsequent analyses. They are not symmetric due to the presence of atmospheric effects. In this work, we only used data from the two large-FoV HE detectors for better statistics, also to avoid systematics related to mixing detectors of different FoVs. 
We repeated the procedure for all public Earth-occultation observations (including parts of those observing targets other than `blank-sky' that were conducted before 2021) and stacked the `Earth-occulted' spectra to arrive at an overall background spectrum. 

The accumulated exposure time of `blank-sky' and `Earth-occulted' observations by HE is 377  and 187 ks respectively. Although the observation direction of LE is exactly the same with HE, the operation of LE requires larger avoidance angles from the Sun and the bright Earth, which leads to shorter GTIs for LE than HE. 

\subsubsection{Background Bias}
\label{Bias of HE} 
Because the flux of charged particles in space varies temporally and spatially, the HE background changes significantly and periodically \citep[see][Figure 4]{liaoBackgroundModelHighenergy2020}. Therefore, the background might not have been the same during the `blank-sky' and `Earth-occulted' periods, which could thus lead to a significant systematic bias in the measurement of DXRB.

We experimented with making use of the HE `blind' detector module to minimize the systematic bias, as it only measures the particle background at all times. We grouped the `blank-sky' and `Earth-occulted' observations for the `blind' detector separately, then sorted them by exposure time in each group. To break the degeneracy between systematic uncertainties due to background variation and due to difference in detector spectral resolution, we chose to focus on a line-free spectral region between 40-45 keV, and computed the count rate in that energy band for each observation.

\begin{figure}
\begin{center}
\includegraphics[width=1.0 \columnwidth]{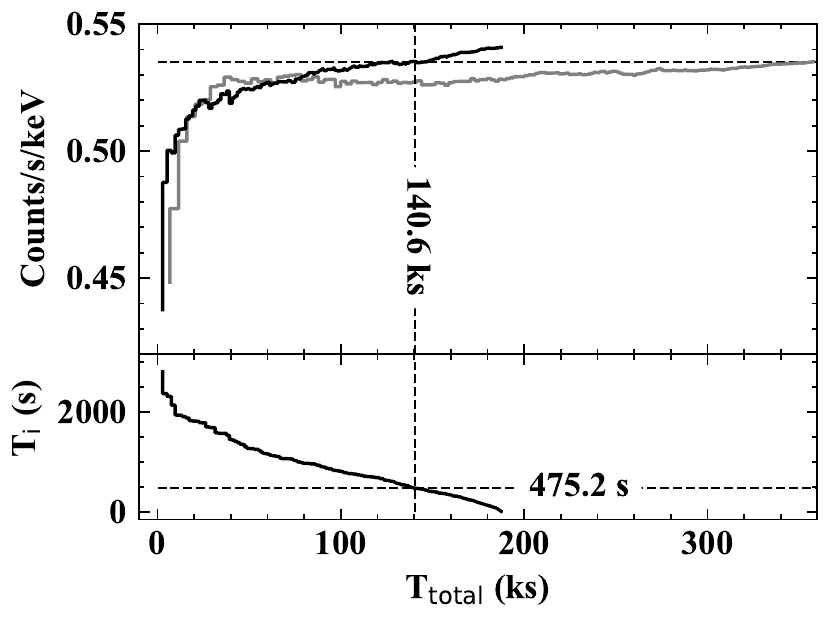}
\end{center}
\caption{({\it top}) Comparison between the average count rates of the `blind' detector for the `blank-sky' period (in grey) and `Earth-occulted' period (in black). The data are cumulatively added from the longest to the shortest exposure time, and the count rates are computed in the 40-45 keV band. The dashed horizontal line indicates the background rate, if all `blank-sky' observations are included. ({\it bottom}) The mininum exposure time $T_i$ of the $i$th observation in the sorted list of `Earth-occulted' observations (from longest to shortest) used for stacking analyses.  } \label{count rate of Det16}
\end{figure}

Fig.~\ref{count rate of Det16} shows the average count rates as a function of cumulative exposure time for the `blank-sky' and `Earth-occulted' periods, respectively. It is quite apparent that, beyond a cumulative exposure time of about 90 ks, the use of `Earth-occulted' observations to assess the background would lead to a significant bias. To address the issue, we empirically determined a threshold on the exposure time to select individual `Earth-occulted' observations for deriving an overall background spectrum. 

\begin{figure}
\begin{center}
\includegraphics[width=1.0 \columnwidth]{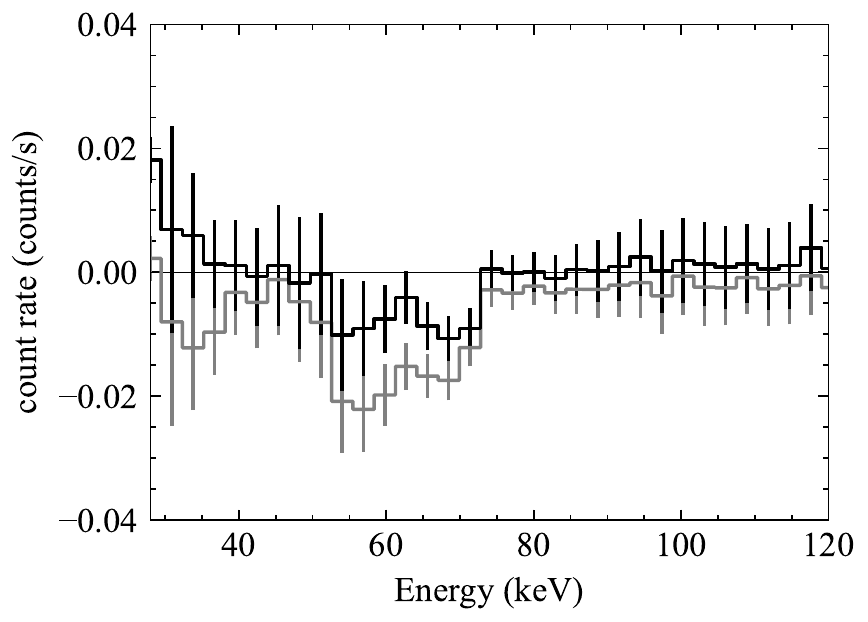}
\end{center}
\caption{HE difference spectrum between `blank-sky' and `Earth-occulted' observations with the `blind' detector. The results are shown for cases with (in black) and without (in grey) bias correction (see the main text), respectively. Note that, because no response matrix is available for the `blind' detector, the response matrices of the two large-FoV detectors are averaged for converting channels to energies here. } \label{redisual of blind}
\end{figure}

As shown in Fig.~\ref{count rate of Det16}, for the `Earth-occulted' observations to reach the same background level as all `blank-sky' observations combined, the threshold is 475.2 s, corresponding to a cumulative exposure time of 140.6 ks. Therefore, we filtered out the `Earth-occulted' observations whose exposure times are shorter than 475.2 s, for constructing the background spectrum. This would remove the background bias in 40-45 keV band. Fig.~\ref{redisual of blind} shows that, with this correction, the systematic bias is much reduced at other energies as well. The remaining bias is present mainly in the energy range with variable instrumental lines (at 31, 57, 65 keV), which were modelled with additional Gaussian components in spectral fitting. 

Applying the procedure above to the normal (large-FoV) detectors, we selected and stacked the `Earth-occulted' spectra, as for the `blind' detector, to derive an overall background spectrum for subsequent DXRB spectral modelling. 

\subsection{Systematic uncertainty}
\label{Systematic uncertainty}

As for remaining systematic uncertainties, we used a bootstrapping method to quantify them. Specifically, for HE, we selected 256 observations, from which the time intervals corresponding to `blank-sky' observations were identified, amounting to a total exposure time of 377 ks. From the 256 `blank-sky' time intervals, we drew samples, allowing repetition, until the total exposure time exceeded 377 ks. The data from the sampled intervals were then stacked, as before, to produce a co-added spectrum with \textit{addspec}. This re-sampling procedure was repeated 10000 times, and also for `Earth-occulted' observations. From the sample spectra, we computed the standard deviation of the count rates in each channel ($\sigma_{\mathrm{total}}$), which contains contributions from systematic uncertainties ($\sigma_{\mathrm{sys}}$) and statistical uncertainties ($\sigma_{\mathrm{stat}}$), with $\sigma_{\mathrm{sys}}=\sqrt{\sigma_{\mathrm{total}}^2-\sigma_{\mathrm{stat}}^2}$. 

\begin{figure}
\begin{center}
\includegraphics[width=1.0 \columnwidth]{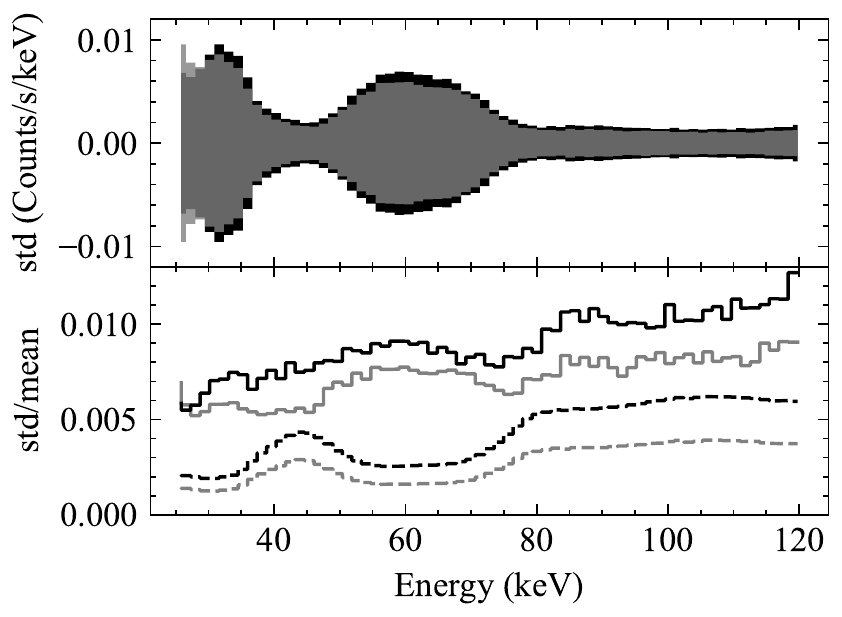}
\end{center}
\caption{HE spectral uncertainties. {\it top}: Standard deviation. {\it bottom}: Fractional standard deviation (solid line) and fractional statistical uncertainty (dashed line). The results are shown in grey and black for the `blank-sky' and `Earth-occulted' spectra, respectively.
} 
\label{pic:syserr}
\end{figure}

We carried out a similar bootstrapping analysis for LE. Our results indicate that systematic uncertainties are quite small for LE, compared with statistical uncertainties, but are quite significant for HE. Fig.~\ref{pic:syserr} shows the results for HE, both in terms of absolute and fractional standard deviations, for the `blank-sky' and `Earth-occulted' observations, respectively. Note the relatively small dispersion in the energy range of 40-45 keV (which we chose as a reference band for minimizing background bias). The statistical uncertainties are larger for the `Earth-occulted' observations than for the `blank-sky' observations, because the accumulated exposure time of `Earth-occulted' observations is shorter.

The uncertainties in the calibration of effective areas and response matrices of the detector modules may also contribute to the systematic uncertainties (\citealt{liInflightCalibrationInsightHard2020}), but they are known to be relatively small and thus neglected in this work.

\section{Spectral Modelling}
\label{Results}
Using the stacked `blank-sky' spectrum and the corresponding background spectrum, we carried out spectral modelling in Sherpa (v4.13.0).
\subsection{LE}
\begin{figure}
\begin{center}
\includegraphics[width=1.0 \columnwidth]{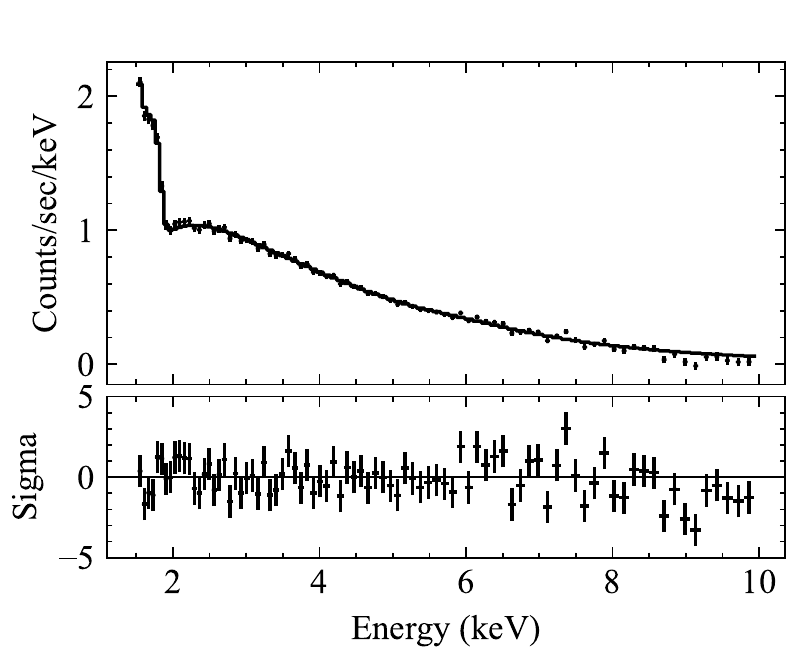}
\end{center}
\caption{The stacked LE spectrum of the `blank-sky' observations. The best-fit power-law model is shown in solid line. The residuals are shown in the bottom panel.} \label{LE_spectra}
\end{figure}

To avoid known calibration issues at both ends of the LE passing band, we chose to limit the energy range to 1.5-10 keV for spectral fitting. Assuming the DXRB contribution is negligible above 10 keV (\citealt{liaoBackgroundModelLowenergy2020}), we computed a scale factor ($backscale=18.0$) by comparing the background spectrum and the `blank-sky' spectrum in the energy range of 10-12.5 keV, which was used to scale the background spectrum in spectral modelling. Fig.~\ref{LE_spectra} shows the results of a fit to the LE spectrum with a simple power-law model. Note that we neglected interstellar absorption, because it hardly affects the spectrum in the chosen energy range.  Overall, the fit is satisfactory. The known instrumental lines at 7.472,  8.041, and 8.631 keV ( \citealt{liInflightCalibrationInsightHard2020}) might be the cause of relatively large residuals at those energies.  

\begin{figure}
\begin{center}
\includegraphics[width=1.0 \columnwidth]{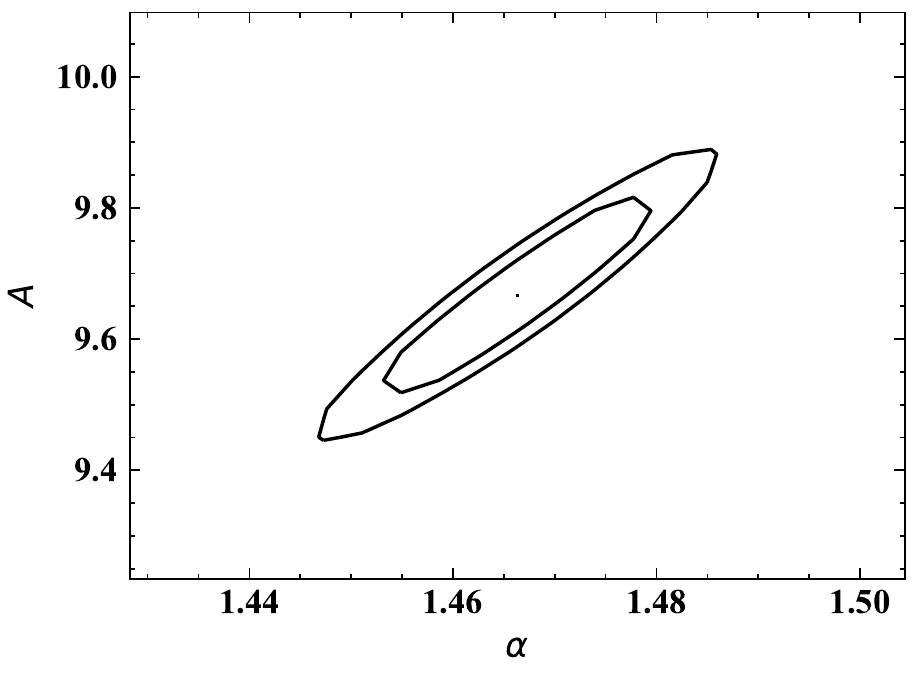}
\end{center}
\caption{Modelling the LE spectrum with a power law. The contours represent the 68\% and 90\% confidence regions of the best-fit parameters.} 
\label{LE_RegionProjection}
\end{figure}

After correcting effective FoV of LE to $9.36~ \rm{arcmin^{2}}$, according to \citealt{nangInorbitCalibrationPointspread2020}, the best-fit photon index and normalization are: $\alpha =1.47_{-0.01}^{+0.01}$  and $A=\rm 9.67_{-0.15}^{+0.15} \rm ~ph~cm^{-2}~s^{-1}~keV^{-1}~sr^{-1}$, respectively, where the uncertainties represent 68\% confidence intervals. The reduced $\chi^2$ of the fit is: $\chi^2_{\nu} \approx 1.3 $ (with 82 degrees of freedom). 
Fig.~\ref{LE_RegionProjection} shows the 2-D confidence regions of the model parameters, indicating that the power-law normalization is strongly correlated with the photon index.

\subsection{HE}

\begin{figure}
\begin{center}
\includegraphics[width=1.0 \columnwidth]{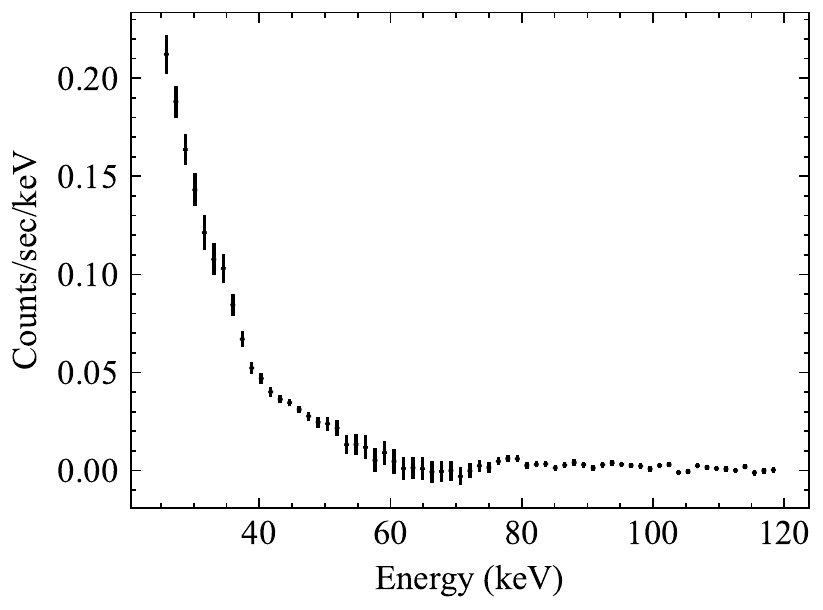}
\end{center}
\caption{The background-subtracted HE spectrum of the `blank-sky' fields. It was made with data from the large-FoV detectors. Systematic uncertainties have been taken into account. } \label{pic:HEspec}
\end{figure}

Fig.~\ref{pic:HEspec} shows the HE spectrum of `blank sky', with the background subtracted. The spectrum was made with data from the two large-FoV detectors only, taking into account systematic uncertainties, because the small-FoV detectors offer lower signal-to-noise ratios in the measurement.

As shown in Sec.~\ref{Bias of HE}, we have effectively removed a systematic bias in the background modelling based on measurements between 40 and 45 keV. Moreover, we quantified systematic uncertainties in the spectrum (see Sec.~\ref{Systematic uncertainty}). It is apparent that between 55-75 keV, where variable instrumental lines are present in the background spectrum, the uncertainties are larger; we had to add a Gaussian component to the model, in order to minimize residuals in that energy range during spectral fitting.

Extrapolating the model that fits the LE spectrum to higher energies, we found that it would lie above the HE data points. It is thus apparent that a spectral roll-over would be required to fit the joint LE-HE spectrum. We implemented the roll-over with the following cutoff power law:

\begin{equation}
F=A~E^{-\alpha}~e^{-E/E_c} \rm ~ph~cm^{-2}~s^{-1}~keV^{-1}~sr^{-1}.
\\
\label{eqn:cutoff_pl}
\end{equation}

\begin{figure}
\begin{center}
\includegraphics[width=1.0 \columnwidth]{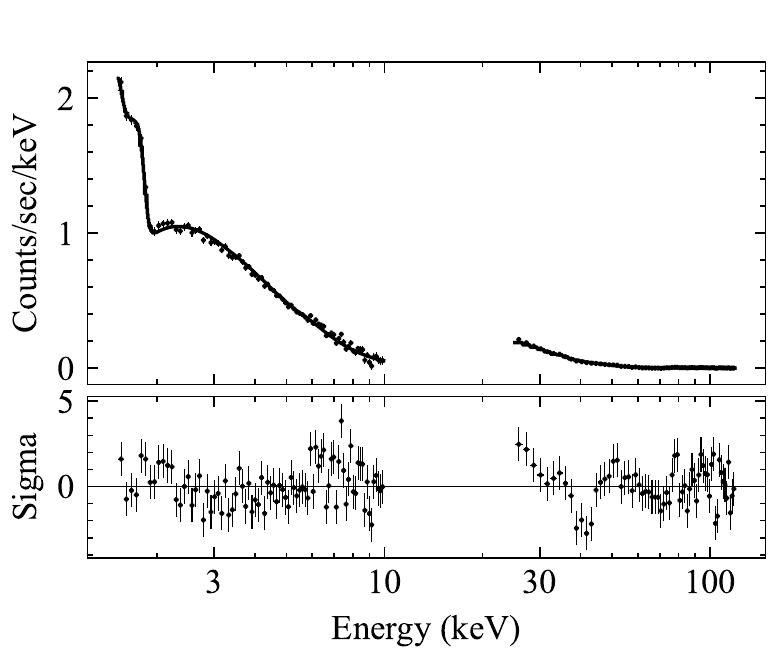}
\end{center}
\caption{Joint LE-HE spectral modelling with a cut-off power-law model. The measured spectra are shown in the top panel, along with the best-fit model (in solid line). The residuals are shown in the bottom panel. } 
\label{pic:joint_fit}
\end{figure}

 Applying the model to the joint spectrum,  we obtained the following best-fit parameters: $A=9.5^{+0.1}_{-0.1}~ \rm  ph~cm^{-2}~s^{-1}~sr^{-1}$, $\alpha=1.33^{+0.01}_{-0.01}$, and $E_c=24^{+1}_{-1}~ \rm  keV$. The reduced $\chi^2$ of the overall fit is about 1.2 (with 143 degrees of freedom). The best-fit model is shown  Fig.~\ref{pic:joint_fit}, along with the residuals.

\section{Discussion}
\label{Discussion}

With the LE data alone, the measured DXRB spectrum follows a simple power law. The measured power-law normalization of the LE spectrum is about $9.67~\rm ph~cm^{-2}~s^{-1}~keV^{-1}~sr^{-1}$ (at 1 keV),  which is in good agreement with the RXTE/PCA result, while the measured photon index (about $1.47$) is softer (\citealt{revnivtsevSpectrumCosmicXray2003}); on the other hand, the normalization is about 11\% lower than the Chandra measurement, while the photon index agrees (\citealt{cappellutiChandraCOSMOSLegacy2017}). Adding the HE data, it becomes apparent that the spectrum rolls over at high energies. A cut-off power law can describe the joint LE-HE spectrum fairly well, with the photon index now about 1.33, which is close to the result of \cite{gruberSpectrumDiffuseCosmic1999b} obtained with HEAO-1, but the cut-off energy is about 24 keV, which appears to be significantly lower than the HEAO-1 result (about 41 keV). 

In the case of HEAO-1 observations, however, the detector background was directly measured, while we had to rely on Earth-occultation observations to infer the background for `blank-sky' observations with HE, which could suffer from a number of systematic uncertainties, as pointed out by \cite{churazovINTEGRALObservationsCosmic2007} in their analyses of the INTEGRAL observations. Most significant are those that directly affect Earth-occultation observations, mainly including the DXRB flux reflected by the Earth atmosphere (referred to as `Earth albedo’ thereinafter) and X-rays emitted by the atmosphere due to cosmic-ray interaction (referred to as `atmospheric emission' thereinafter).

Based on simulation results, \cite{churazovEarthXrayAlbedo2008} found that the spectrum shape of the Earth albedo does not depend strongly on the altitude of low-earth orbits, but does on photon energy. We used an empirical formula that they derived to describe the Earth albedo:
\begin{eqnarray}
A(E)=\frac{1.22}
{\left(\frac{E}{28.5}\right)^{-2.54}+\left(\frac{E}{51.3}\right)^{1.57}-0.37}\times
\nonumber \\
\frac{2.93+\left(\frac{E}{3.08}\right)^4}{1+\left(\frac{E}{3.08}\right)^4}\times
\nonumber \\
\frac{0.123+\left(\frac{E}{91.83}\right)^{3.44}}{1+\left(\frac{E}{91.83}\right)^{3.44}},
\label{eqn:refl}
\end{eqnarray}
It should be noted, however, that the reflected emission depends on the viewing angle (see Figure 5 of \citealt{churazovEarthXrayAlbedo2008}): it is much stronger when looking toward the Earth limb than toward the centre of the Earth.
Eq.~\ref{eqn:refl} is the angle-averaged Earth albedo for an instrument whose FoV covers the whole Earth (e.g., INTEGRAL, SWIFT/BAT, etc.). Because the FoV of HE is small (even for the large-FoV detectors), we multiplied the results from  Eq.~\ref{eqn:refl} with an additional factor $A_{\mathrm{albedo}}$. 

As for the atmospheric emission, \cite{sazonovHardXrayEmission2007} conducted Monte Carlo simulations with GEANT4 on the interaction between cosmic rays and the atmosphere, and found that the results are not sensitive to the cosmic-ray spectrum. The resulting atmospheric emission can be described by: 
\begin{equation}
S(E)=\frac{C}{(E/44)^{-5}+(E/44)^{1.4}}  \rm ~ph~cm^{-2}~s^{-1}~keV^{-1}~sr^{-1} ,
\label{eq:atm_fit}
\end{equation}
where $C$ depends on solar modulation level, geomagnetic cut-off rigidity and viewing angle. 
Taking into account this effect, as well as the Earth albedo, we derived the corrected DXRB flux as $F(E) \cdot (1-A'(E)) - S(E)$, where $A'(E) = A_{\mathrm{albedo}} \times A(E)$, where $F(E)$ is the spectral model for DXRB.  

For a direct comparison with the HEAO-1 result, we replaced the cutoff power-law model with an empirical model proposed by \cite{gruberSpectrumDiffuseCosmic1999b}: 
\begin{equation}
F_{\rm gruber}=\left\{
\begin{array}{ll}
7.877 E^{-1.29} e^{-E / 41.13}~~~~\mathrm{~\frac{ph}{cm^2~s~keV~sr}} & {3-60 \mathrm{keV}} \\
(0.0259\left(\frac{E}{60}\right)^{-5.5} \\
\qquad +0.504\left(\frac{E}{60}\right)^{-1.58}~~~~~\mathrm{\frac{ph}{cm^{2}~s~keV~sr}}\\  
\qquad +0.0288\left(\frac{E}{60}\right)^{-1.05})/E   & {>60 \mathrm{keV}},
\end{array} \right.
\label{eqn:gruber}
\end{equation}
which represents the best fit to the DXRB spectrum as measured with HEAO-1. To see whether the model can also describe the shape of the HE spectrum in our case, we multiplied $F_{\rm gruber}$ with an adjustable factor ($A_{\rm gruber}$). Here, we also needed to correct for the effects of the Earth albedo and atmospheric emission: $F'_{\rm gruber}  \cdot (1-A'(E)) - S(E) $, where the $F'_{\rm gruber} = A_{\rm gruber} \times F_{\rm gruber}$.

\begin{figure}
\begin{center}
\includegraphics[width=1.0 \columnwidth]{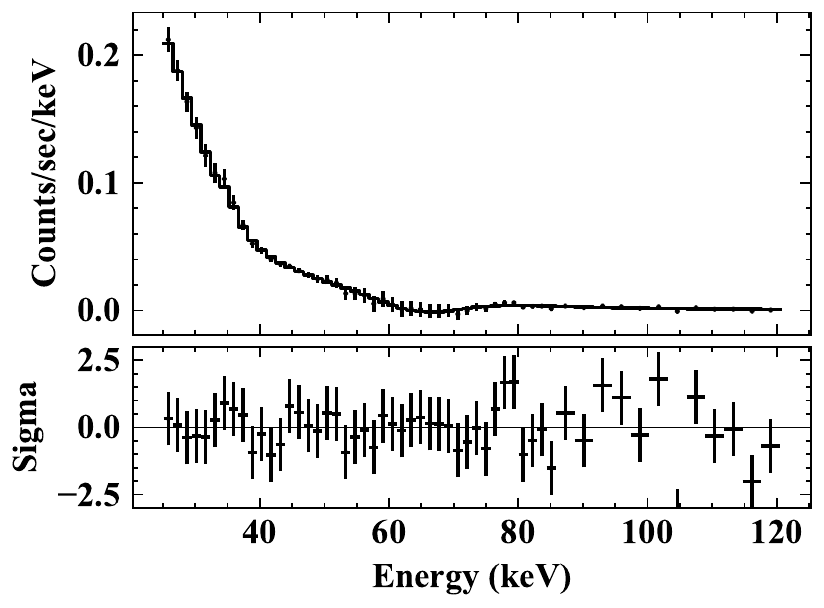}
\end{center}
\caption{Modelling of the HE spectrum.  The upper panel shows the best-fit model (in solid line), and the bottom panel shows the residuals.} 
\label{pic:gruber}
\end{figure}

\begin{figure}
\begin{center}
\includegraphics[width=1.0 \columnwidth]{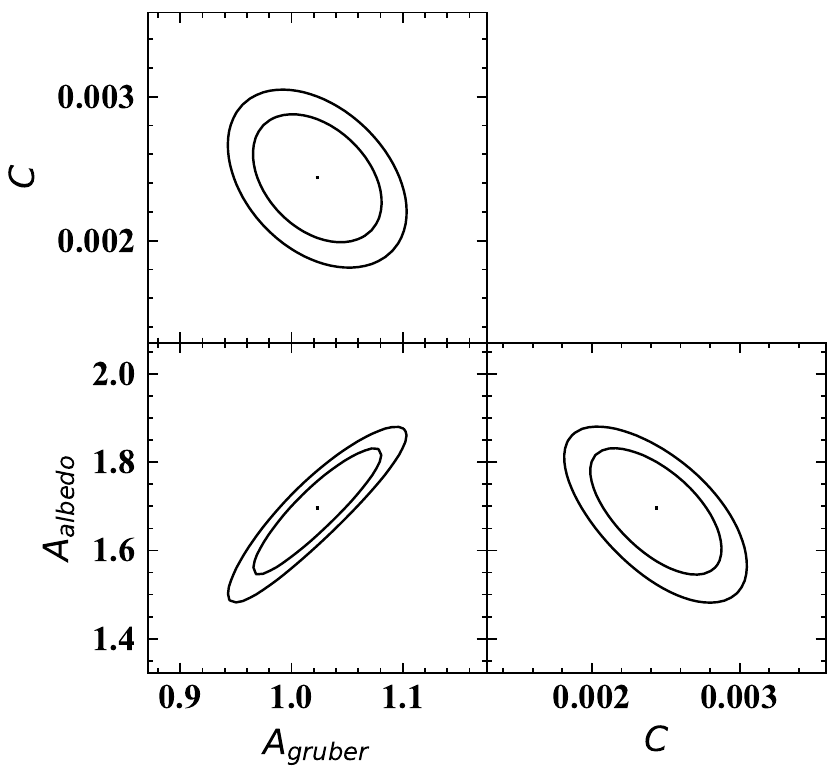}
\end{center}
\caption{The 68\% and 90\% confidence regions of the best-fit model parameters, as shown in Fig.~\ref{pic:gruber}. } \label{pic:gruber_confidence}
\end{figure}

Fitting the HE spectrum with this model, we arrived at the best-fit parameters: $A_{\rm gruber}=1.02^{+0.06}_{-0.06}$, $A_{\rm albedo}=1.70^{+0.14}_{-0.14}$ and $C = 2.4^{+0.4}_{-0.4}\times10^{-3}~\mathrm{ph~cm^{-2}~s^{-1}~keV^{-1}~sr^{-1}}$, indicating the the Grubber model is also preferred by the HE data. The derived Earth albedo is 70\% and 20\% higher than the value measured by \citealt{churazovEarthXrayAlbedo2008} and \citealt{2010A&A...512A..49T}, respectively based on INTEGRAL observations, while the atmospheric emission is much lower than their measurements. The orbits of the Insight-HXMT and INTEGRAL satellites are similar, so the discrepancies are likely attributable to the difference in the FoVs of the instruments, because the Earth albedo is known to be limb-brightened (\citealt{churazovEarthXrayAlbedo2008}), and the atmospheric emission centrally-brightened (\citealt{sazonovHardXrayEmission2007}). Based on the distribution of viewing angles of our selected observations, we find that the `Earth-occultation' observations are weighted more towards the Earth limb in our case, compared to whole-Earth coverage with INTEGRAL, which may explain the measured $A_{\rm albedo}$ and $C$. 
The atmosphere emission also depends on solar modulation level and geomagnetic cutoff rigidity. The solar modulation level during our observations is similar to that during the INTEGRAL observations (\citealt{churazovEarthXrayAlbedo2008},\citealt{sazonovHardXrayEmission2007}), based on the number of sunspots. On the other hand, our data filtering criteria favour large geomagnetic cut-off rigidity (COR > 8), which might also partly explain the lower level of the atmospheric emission (although we do not know the geomagnetic cut-off rigidity range associated with the INTEGRAL observations.) The best-fit model is shown in Fig.~\ref{pic:gruber}, while Fig.~\ref{pic:gruber_confidence} shows the 2-D confidence regions of model parameters.

\begin{figure}
\begin{center}
\includegraphics[width=1.0 \columnwidth]{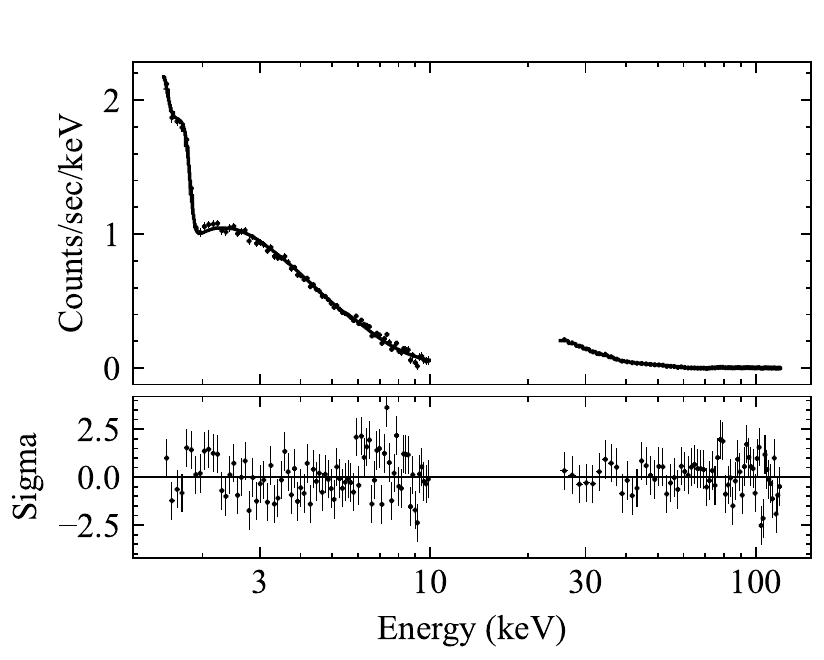}
\end{center}
\caption{As for Fig.~\ref{pic:joint_fit}, but with the effects of Earth albedo and atmospheric emission considered. Note the significant improvement in minimizing the residuals (see the bottom panel). } 
\label{pic:joint_fit_albedo_atm}
\end{figure}

\begin{figure}
\begin{center}
\includegraphics[width=1.0 \columnwidth]{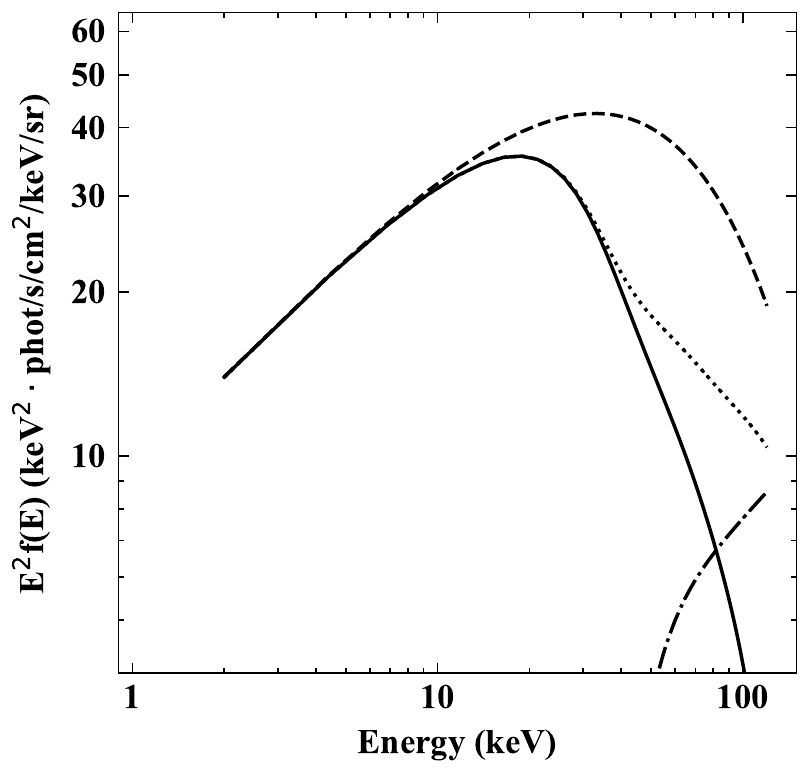}
\end{center}
\caption{The spectral energy distributions derived from the blank-sky observations (in solid line), based on the best-fit cut-off power law as shown in Fig. \ref{pic:joint_fit_albedo_atm}. The SED of the DXRB is shown in dashed line, after the effects of the Earth albedo and atmospheric emission both considered. Also shown are the SED with only the effects of the Earth albedo considered (in dotted line), and the atmospheric emission (in dot-dashed line), for comparison.}
\label{pic:model_eeplot}
\end{figure}

\begin{figure}
\begin{center}
\includegraphics[width=1.0 \columnwidth]{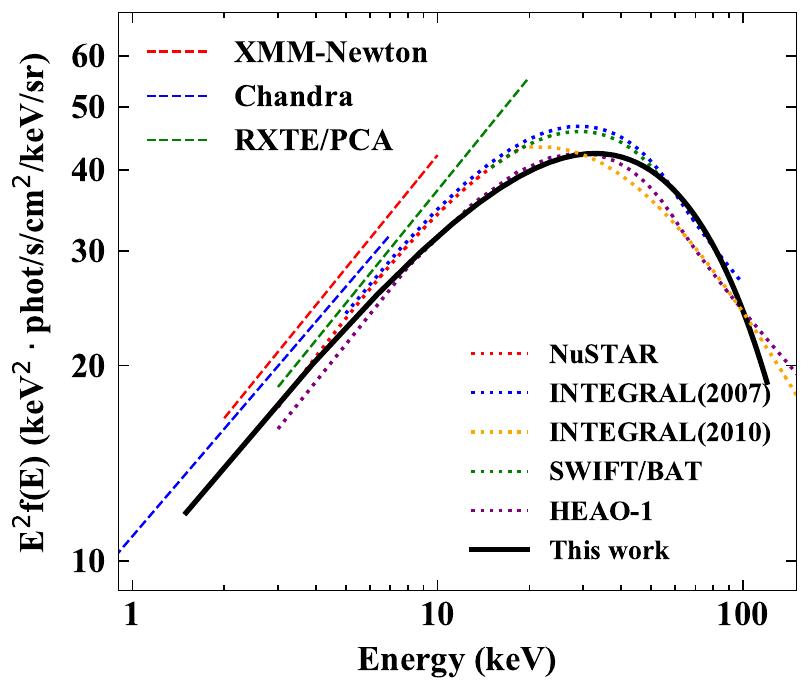}
\end{center}
\caption{The spectral energy distribution of the diffuse X-ray background. The results of this work are shown in black solid lines. For comparison, the results obtained with XMM-Newton (\citealt{lumbXrayBackgroundMeasurements2002}), Chandra (\citealt{cappellutiChandraCOSMOSLegacy2017}), RXTE/PCA (\citealt{revnivtsevSpectrumCosmicXray2003}) are shown in red, blue, green dashed lines, respectively, at lower energies, and the results obtained with NuSTAR (\citealt{krivonosNuSTARMeasurementCosmic2021}), INTEGRAL (including both (\citealt{churazovINTEGRALObservationsCosmic2007} and \citealt{2010A&A...512A..49T}), SWIFT/BAT (\citealt{ajelloCosmicRayBackground2008}), HEAO-1 (\citealt{gruberSpectrumDiffuseCosmic1999b}) are shown in red, blue, orange, green, purple dotted lines, respectively, at higher energies.
} \label{CXB_model_HXMT} 
\end{figure}

Because the background was directly measured with the `blind' detectors in the case of LE, we did not need to consider the effects of the Earth albedo and atmospheric emission (which are also expected to be negligible in the LE passing band). In order to fit the LE and HE spectra jointly, we replaced the Gruber model (Eq.~\ref{eqn:gruber}) with a cut-off power-law function (Eq.~\ref{eqn:cutoff_pl}), but with $A_{\mathrm{albedo}}$ and $C$ fixed at the above best-fit values (see Fig.~\ref{pic:joint_fit_albedo_atm}) . Now, the best-fit parameters are:   $A=9.57^{+0.13}_{-0.13}~\rm ph~cm^{-2}~s^{-1}~keV^{-1}~sr^{-1}$, $\alpha = 1.402^{+0.015}_{-0.015}$, $E_c=55^{+4}_{-4} ~\rm keV$. The reduced $\chi^2$ of the fit is about 1.0. Based on this best-fit model, Fig. \ref{pic:model_eeplot} shows the derived spectral energy distribution (SED) of the DXRB, after considering the effects of the Earth albedo and atmospheric emission. 
Fig.~\ref{CXB_model_HXMT} shows a comparison between our result and the published SEDs in the literature. Our measured SED is in good agreement with the HEAO-1 results (\citealt{1980ApJ...235....4M}; \citealt{gruberSpectrumDiffuseCosmic1999b}), especially between 6 and 30 keV, but is about 7\% lower than the INTEGRAL result reported by \cite{churazovINTEGRALObservationsCosmic2007} and 5\% lower than the SWIFT/BAT result (\citealt{ajelloCosmicRayBackground2008}). At higher energies, our SED shows a sharper spectral roll-over than the HEAO-1 SED, although this is quite uncertain due to the model-dependent nature of spectral unfolding process. We note that the HEAO-1 SED is also favoured by \citealt{2010A&A...512A..49T}, based on their INTEGRAL observations (see, however, \cite{churazovINTEGRALObservationsCosmic2007}).

At energies above about 1 keV, the DXRB is generally thought to mostly originate in integrated emission from AGN. Chandra directly resolved about 91\% of their measured DXRB flux (below 7 keV) into discrete sources, including not only those it detected but sources detected in other wave bands, the majority of which are AGN (\citealt{cappellutiChandraCOSMOSLegacy2017}). In hard X-rays, NuSTAR possesses excellent angular resolution, and the AGN that it detected can account for about 33-39\% of the DXRB flux between 8-24 keV, depending on the assumed normalization of the DXRB spectrum (\citealt{2016ApJ...831..185H}); the high value (39\%) is also favored by the results of our work.
Going further up in energy, between 20-60 keV, it becomes difficult to resolve contribution from individual sources due to inadequate angular resolution (e.g., \citealt{2021NewAR..9201612K}), as is the case in the present work.  
INTEGRAL discovered a significant population of Compton-thick AGN (e.g., \citealt{2007A&A...462...57S},\citealt{2008A&A...487..509S}), and solved the long-standing puzzle in the mismatch between the shape of the DXRB spectrum and that of the integrated AGN spectrum. If DXRB is entirely attributable to AGN in hard X-rays, using AGN population synthesis models, one could constrain the AGN number density and fractions of different types of AGN. Through modelling the DXRB spectrum measured with ASCA (\citealt{1995PASJ...47L...5G}), Chandra (\citealt{cappellutiChandraCOSMOSLegacy2017}), RXTE/PCA (\citealt{revnivtsevSpectrumCosmicXray2003}) and SWIFT/BAT (\citealt{ajelloCosmicRayBackground2008}), \cite{2019ApJ...871..240A} concluded that $50\pm 9\% (56\pm 7\%)$  of the AGN within z$\simeq$0.1 (1.0) are Compton-thick.
In this work, the derived normalization of the DXRB spectrum is about 8\% lower, so our results would imply a lower number density of Compton-thick AGN. 

From the programmatic point of view, our results show that the LE and HE aboard Insight-HXMT have been calibrated sufficiently well for us to attempt such a difficult measurement of the DXRB. Improving background calibration was one of the initial motivations for this work. The issues are summarized as follows:
\begin{enumerate}
\item Earth-occulted observations appear to suggest a light-leak issue for LE which might be linked to auroral emission as detected for instance by INTEGRAL in more recent Earth observations conducted during periods of stronger solar activity. (\citealt{2021NewAR..9301616T}). We were not able to use those observations to assess detector background.
\item For HE, the difference in energy resolution among the detector modules has prevented us from using the `blind' detector to measure the particle background. The issue is particular serious in the energy range of 55-75 keV, where the particle-induced spectral lines are relatively strong and variable.
\end{enumerate}

We do not expect that adding data from future observations would improve the result, because systematic uncertainties are already dominating, not to mention the fact the performance of detectors is also degrading.   
On the other hand, we might be able to improve the HE background model based on data from the blind detector alone. That would allow us to carry out analyses like in the case of HEAO-1, without having to consider the effects of the Earth albedo and the atmospheric emission. This possibility will be investigated in the future.

\section*{Acknowledgements}

We thank the anonymous referee for the helpful comments that have improved the manuscript. This work was supported in part by the National Natural Science Foundation of China (NSFC) through Grant No. U1838107. We also acknowledge additional support from the National Program on Key Research and Development Project through grant 2021YFA0718500 and from the NSFC through Grants U1838202 and U1838201. The work made use of data from the Insight-HXMT mission, a project funded by the China National Space Administration (CNSA) and the Chinese Academy of Sciences (CAS). The authors acknowledge the Tsinghua Astrophysics High-Performance Computing platform at Tsinghua University for providing computational and data storage resources that have contributed to the research results reported within this paper.

\section*{Data Availability}
The data that support the findings of this study are available from Insight-HXMT’s data archive (http://enghxmt.ihep.ac.cn).



\bibliographystyle{mnras}
\bibliography{DXRB} 

\begin{thebibliography}{}
\makeatletter
\relax
\def\mn@urlcharsother{\let\do\@makeother \do\$\do\&\do\#\do\^\do\_\do\%\do\~}
\def\mn@doi{\begingroup\mn@urlcharsother \@ifnextchar [ {\mn@doi@}
  {\mn@doi@[]}}
\def\mn@doi@[#1]#2{\def\@tempa{#1}\ifx\@tempa\@empty \href
  {http://dx.doi.org/#2} {doi:#2}\else \href {http://dx.doi.org/#2} {#1}\fi
  \endgroup}
\def\mn@eprint#1#2{\mn@eprint@#1:#2::\@nil}
\def\mn@eprint@arXiv#1{\href {http://arxiv.org/abs/#1} {{\tt arXiv:#1}}}
\def\mn@eprint@dblp#1{\href {http://dblp.uni-trier.de/rec/bibtex/#1.xml}
  {dblp:#1}}
\def\mn@eprint@#1:#2:#3:#4\@nil{\def\@tempa {#1}\def\@tempb {#2}\def\@tempc
  {#3}\ifx \@tempc \@empty \let \@tempc \@tempb \let \@tempb \@tempa \fi \ifx
  \@tempb \@empty \def\@tempb {arXiv}\fi \@ifundefined
  {mn@eprint@\@tempb}{\@tempb:\@tempc}{\expandafter \expandafter \csname
  mn@eprint@\@tempb\endcsname \expandafter{\@tempc}}}

\bibitem[\protect\citeauthoryear{{Ajello} et~al.,}{{Ajello}
  et~al.}{2008}]{ajelloCosmicRayBackground2008}
{Ajello} M.,  et~al., 2008, \mn@doi [\apj] {10.1086/592595}, \href
  {https://ui.adsabs.harvard.edu/abs/2008ApJ...689..666A} {689, 666}

\bibitem[\protect\citeauthoryear{{Ananna} et~al.,}{{Ananna}
  et~al.}{2019}]{2019ApJ...871..240A}
{Ananna} T.~T.,  et~al., 2019, \mn@doi [\apj] {10.3847/1538-4357/aafb77}, \href
  {https://ui.adsabs.harvard.edu/abs/2019ApJ...871..240A} {871, 240}

\bibitem[\protect\citeauthoryear{{Cappelluti} et~al.,}{{Cappelluti}
  et~al.}{2017}]{cappellutiChandraCOSMOSLegacy2017}
{Cappelluti} N.,  et~al., 2017, \mn@doi [\apj] {10.3847/1538-4357/aa5ea4},
  \href {https://ui.adsabs.harvard.edu/abs/2017ApJ...837...19C} {837, 19}

\bibitem[\protect\citeauthoryear{{Chen} et~al.,}{{Chen}
  et~al.}{2020}]{chenLowEnergyXray2020}
{Chen} Y.,  et~al., 2020, \mn@doi [Science China Physics, Mechanics, and
  Astronomy] {10.1007/s11433-019-1469-5}, \href
  {https://ui.adsabs.harvard.edu/abs/2020SCPMA..6349505C} {63, 249505}

\bibitem[\protect\citeauthoryear{{Churazov} et~al.,}{{Churazov}
  et~al.}{2007}]{churazovINTEGRALObservationsCosmic2007}
{Churazov} E.,  et~al., 2007, \mn@doi [\aap] {10.1051/0004-6361:20066230},
  \href {https://ui.adsabs.harvard.edu/abs/2007A&A...467..529C} {467, 529}

\bibitem[\protect\citeauthoryear{{Churazov}, {Sazonov}, {Sunyaev}  \&
  {Revnivtsev}}{{Churazov} et~al.}{2008}]{churazovEarthXrayAlbedo2008}
{Churazov} E.,  {Sazonov} S.,  {Sunyaev} R.,   {Revnivtsev} M.,  2008, \mn@doi
  [\mnras] {10.1111/j.1365-2966.2008.12918.x}, \href
  {https://ui.adsabs.harvard.edu/abs/2008MNRAS.385..719C} {385, 719}

\bibitem[\protect\citeauthoryear{{Croft}, {Di Matteo}, {Dav{\'e}}, {Hernquist},
  {Katz}, {Fardal}  \& {Weinberg}}{{Croft}
  et~al.}{2001}]{croftHydrodynamicSimulationCosmological2001}
{Croft} R. A.~C.,  {Di Matteo} T.,  {Dav{\'e}} R.,  {Hernquist} L.,  {Katz} N.,
   {Fardal} M.~A.,   {Weinberg} D.~H.,  2001, \mn@doi [\apj] {10.1086/321632},
  \href {https://ui.adsabs.harvard.edu/abs/2001ApJ...557...67C} {557, 67}

\bibitem[\protect\citeauthoryear{{Galeazzi} et~al.,}{{Galeazzi}
  et~al.}{2011}]{galeazziDXLSoundingRocket2011}
{Galeazzi} M.,  et~al., 2011, \mn@doi [Experimental Astronomy]
  {10.1007/s10686-011-9249-y}, \href
  {https://ui.adsabs.harvard.edu/abs/2011ExA....32...83G} {32, 83}

\bibitem[\protect\citeauthoryear{{Gendreau} et~al.,}{{Gendreau}
  et~al.}{1995}]{1995PASJ...47L...5G}
{Gendreau} K.~C.,  et~al., 1995, \pasj, \href
  {https://ui.adsabs.harvard.edu/abs/1995PASJ...47L...5G} {47, L5}

\bibitem[\protect\citeauthoryear{{Giacconi}, {Gursky}, {Paolini}  \&
  {Rossi}}{{Giacconi} et~al.}{1962}]{giacconiEVIDENCERAYSSOURCES1962}
{Giacconi} R.,  {Gursky} H.,  {Paolini} F.~R.,   {Rossi} B.~B.,  1962,
  Technical Report~11, {Evidence for x Rays From Sources Outside the Solar
  System}, \mn@doi{10.1103/PhysRevLett.9.439.
}

\bibitem[\protect\citeauthoryear{{Gruber}, {Matteson}, {Peterson}  \&
  {Jung}}{{Gruber} et~al.}{1999}]{gruberSpectrumDiffuseCosmic1999b}
{Gruber} D.~E.,  {Matteson} J.~L.,  {Peterson} L.~E.,   {Jung} G.~V.,  1999,
  \mn@doi [\apj] {10.1086/307450}, \href
  {https://ui.adsabs.harvard.edu/abs/1999ApJ...520..124G} {520, 124}

\bibitem[\protect\citeauthoryear{{Guan} et~al.,}{{Guan}
  et~al.}{2021}]{guanPhysicalOriginNonphysical2021}
{Guan} J.,  et~al., 2021, \mn@doi [\mnras] {10.1093/mnras/stab945}, \href
  {https://ui.adsabs.harvard.edu/abs/2021MNRAS.504.2168G} {504, 2168}

\bibitem[\protect\citeauthoryear{{Harrison} et~al.,}{{Harrison}
  et~al.}{2016}]{2016ApJ...831..185H}
{Harrison} F.~A.,  et~al., 2016, \mn@doi [\apj] {10.3847/0004-637X/831/2/185},
  \href {https://ui.adsabs.harvard.edu/abs/2016ApJ...831..185H} {831, 185}

\bibitem[\protect\citeauthoryear{{Hasinger}, {Burg}, {Giacconi}, {Hartner},
  {Schmidt}, {Trumper}  \& {Zamorani}}{{Hasinger}
  et~al.}{1993}]{hasingerXray1993}
{Hasinger} G.,  {Burg} R.,  {Giacconi} R.,  {Hartner} G.,  {Schmidt} M.,
  {Trumper} J.,   {Zamorani} G.,  1993, \aap, \href
  {https://ui.adsabs.harvard.edu/abs/1993A&A...275....1H} {275, 1}

\bibitem[\protect\citeauthoryear{{Krivonos} et~al.,}{{Krivonos}
  et~al.}{2021a}]{2021NewAR..9201612K}
{Krivonos} R.~A.,  et~al., 2021a, \mn@doi [\nar] {10.1016/j.newar.2021.101612},
  \href {https://ui.adsabs.harvard.edu/abs/2021NewAR..9201612K} {92, 101612}

\bibitem[\protect\citeauthoryear{{Krivonos}, {Wik}, {Grefenstette}, {Madsen},
  {Perez}, {Rossland}, {Sazonov}  \& {Zoglauer}}{{Krivonos}
  et~al.}{2021b}]{krivonosNuSTARMeasurementCosmic2021}
{Krivonos} R.,  {Wik} D.,  {Grefenstette} B.,  {Madsen} K.,  {Perez} K.,
  {Rossland} S.,  {Sazonov} S.,   {Zoglauer} A.,  2021b, \mn@doi [\mnras]
  {10.1093/mnras/stab209}, \href
  {https://ui.adsabs.harvard.edu/abs/2021MNRAS.502.3966K} {502, 3966}

\bibitem[\protect\citeauthoryear{{Li}}{{Li}}{2007}]{liHXMTChineseHighenergy2007}
{Li} T.-P.,  2007, \mn@doi [Nuclear Physics B Proceedings Supplements]
  {10.1016/j.nuclphysbps.2006.12.070}, \href
  {https://ui.adsabs.harvard.edu/abs/2007NuPhS.166..131L} {166, 131}

\bibitem[\protect\citeauthoryear{{Li} et~al.,}{{Li}
  et~al.}{2020}]{liInflightCalibrationInsightHard2020}
{Li} X.,  et~al., 2020, \mn@doi [Journal of High Energy Astrophysics]
  {10.1016/j.jheap.2020.02.009}, \href
  {https://ui.adsabs.harvard.edu/abs/2020JHEAp..27...64L} {27, 64}

\bibitem[\protect\citeauthoryear{{Liao} et~al.,}{{Liao}
  et~al.}{2020a}]{liaoBackgroundModelHighenergy2020}
{Liao} J.-Y.,  et~al., 2020a, \mn@doi [Journal of High Energy Astrophysics]
  {10.1016/j.jheap.2020.04.002}, \href
  {https://ui.adsabs.harvard.edu/abs/2020JHEAp..27...14L} {27, 14}

\bibitem[\protect\citeauthoryear{{Liao} et~al.,}{{Liao}
  et~al.}{2020b}]{liaoBackgroundModelLowenergy2020}
{Liao} J.-Y.,  et~al., 2020b, \mn@doi [Journal of High Energy Astrophysics]
  {10.1016/j.jheap.2020.02.010}, \href
  {https://ui.adsabs.harvard.edu/abs/2020JHEAp..27...24L} {27, 24}

\bibitem[\protect\citeauthoryear{{Liu} et~al.,}{{Liu}
  et~al.}{2020}]{liuHighEnergyXray2020}
{Liu} C.,  et~al., 2020, \mn@doi [Science China Physics, Mechanics, and
  Astronomy] {10.1007/s11433-019-1486-x}, \href
  {https://ui.adsabs.harvard.edu/abs/2020SCPMA..6349503L} {63, 249503}

\bibitem[\protect\citeauthoryear{{Lumb}, {Warwick}, {Page}  \& {De
  Luca}}{{Lumb} et~al.}{2002}]{lumbXrayBackgroundMeasurements2002}
{Lumb} D.~H.,  {Warwick} R.~S.,  {Page} M.,   {De Luca} A.,  2002, \mn@doi
  [\aap] {10.1051/0004-6361:20020531}, \href
  {https://ui.adsabs.harvard.edu/abs/2002A&A...389...93L} {389, 93}

\bibitem[\protect\citeauthoryear{{Marshall}, {Boldt}, {Holt}, {Miller},
  {Mushotzky}, {Rose}, {Rothschild}  \& {Serlemitsos}}{{Marshall}
  et~al.}{1980}]{1980ApJ...235....4M}
{Marshall} F.~E.,  {Boldt} E.~A.,  {Holt} S.~S.,  {Miller} R.~B.,  {Mushotzky}
  R.~F.,  {Rose} L.~A.,  {Rothschild} R.~E.,   {Serlemitsos} P.~J.,  1980,
  \mn@doi [\apj] {10.1086/157601}, \href
  {https://ui.adsabs.harvard.edu/abs/1980ApJ...235....4M} {235, 4}

\bibitem[\protect\citeauthoryear{{McCammon} \& {Sanders}}{{McCammon} \&
  {Sanders}}{1990}]{mccammonSoftXrayBackground1990}
{McCammon} D.,  {Sanders} W.~T.,  1990, \mn@doi [\araa]
  {10.1146/annurev.aa.28.090190.003301}, \href
  {https://ui.adsabs.harvard.edu/abs/1990ARA&A..28..657M} {28, 657}

\bibitem[\protect\citeauthoryear{{Nang} et~al.,}{{Nang}
  et~al.}{2020}]{nangInorbitCalibrationPointspread2020}
{Nang} Y.,  et~al., 2020, \mn@doi [Journal of High Energy Astrophysics]
  {10.1016/j.jheap.2020.01.002}, \href
  {https://ui.adsabs.harvard.edu/abs/2020JHEAp..25...39N} {25, 39}

\bibitem[\protect\citeauthoryear{{Revnivtsev}, {Gilfanov}, {Sunyaev}, {Jahoda}
  \& {Markwardt}}{{Revnivtsev} et~al.}{2003}]{revnivtsevSpectrumCosmicXray2003}
{Revnivtsev} M.,  {Gilfanov} M.,  {Sunyaev} R.,  {Jahoda} K.,   {Markwardt} C.,
   2003, \mn@doi [\aap] {10.1051/0004-6361:20031386}, \href
  {https://ui.adsabs.harvard.edu/abs/2003A&A...411..329R} {411, 329}

\bibitem[\protect\citeauthoryear{{Sazonov}, {Churazov}, {Sunyaev}  \&
  {Revnivtsev}}{{Sazonov} et~al.}{2007a}]{sazonovHardXrayEmission2007}
{Sazonov} S.,  {Churazov} E.,  {Sunyaev} R.,   {Revnivtsev} M.,  2007a, \mn@doi
  [\mnras] {10.1111/j.1365-2966.2007.11746.x}, \href
  {https://ui.adsabs.harvard.edu/abs/2007MNRAS.377.1726S} {377, 1726}

\bibitem[\protect\citeauthoryear{{Sazonov}, {Revnivtsev}, {Krivonos},
  {Churazov}  \& {Sunyaev}}{{Sazonov} et~al.}{2007b}]{2007A&A...462...57S}
{Sazonov} S.,  {Revnivtsev} M.,  {Krivonos} R.,  {Churazov} E.,   {Sunyaev} R.,
   2007b, \mn@doi [\aap] {10.1051/0004-6361:20066277}, \href
  {https://ui.adsabs.harvard.edu/abs/2007A&A...462...57S} {462, 57}

\bibitem[\protect\citeauthoryear{{Sazonov}, {Revnivtsev}, {Burenin},
  {Churazov}, {Sunyaev}, {Forman}  \& {Murray}}{{Sazonov}
  et~al.}{2008}]{2008A&A...487..509S}
{Sazonov} S.,  {Revnivtsev} M.,  {Burenin} R.,  {Churazov} E.,  {Sunyaev} R.,
  {Forman} W.~R.,   {Murray} S.~S.,  2008, \mn@doi [\aap]
  {10.1051/0004-6361:200809528}, \href
  {https://ui.adsabs.harvard.edu/abs/2008A&A...487..509S} {487, 509}

\bibitem[\protect\citeauthoryear{{Snowden}}{{Snowden}}{1993}]{snowdenImplicationsROSATObservations1993}
{Snowden} S.~L.,  1993, \mn@doi [Advances in Space Research]
  {10.1016/0273-1177(93)90102-H}, \href
  {https://ui.adsabs.harvard.edu/abs/1993AdSpR..13l.103S} {13, 103}

\bibitem[\protect\citeauthoryear{{Snowden}, {Collier}  \& {Kuntz}}{{Snowden}
  et~al.}{2004}]{snowdenXMMNewtonObservationSolar2004}
{Snowden} S.~L.,  {Collier} M.~R.,   {Kuntz} K.~D.,  2004, \mn@doi [\apj]
  {10.1086/421841}, \href
  {https://ui.adsabs.harvard.edu/abs/2004ApJ...610.1182S} {610, 1182}

\bibitem[\protect\citeauthoryear{{T{\"u}rler}, {Chernyakova}, {Courvoisier},
  {Lubi{\'n}ski}, {Neronov}, {Produit}  \& {Walter}}{{T{\"u}rler}
  et~al.}{2010}]{2010A&A...512A..49T}
{T{\"u}rler} M.,  {Chernyakova} M.,  {Courvoisier} T.~J.~L.,  {Lubi{\'n}ski}
  P.,  {Neronov} A.,  {Produit} N.,   {Walter} R.,  2010, \mn@doi [\aap]
  {10.1051/0004-6361/200913072}, \href
  {https://ui.adsabs.harvard.edu/abs/2010A&A...512A..49T} {512, A49}

\bibitem[\protect\citeauthoryear{{T{\"u}rler}, {Tatischeff}, {Beckmann}  \&
  {Churazov}}{{T{\"u}rler} et~al.}{2021}]{2021NewAR..9301616T}
{T{\"u}rler} M.,  {Tatischeff} V.,  {Beckmann} V.,   {Churazov} E.,  2021,
  \mn@doi [\nar] {10.1016/j.newar.2021.101616}, \href
  {https://ui.adsabs.harvard.edu/abs/2021NewAR..9301616T} {93, 101616}

\bibitem[\protect\citeauthoryear{{Zhang}, {Lu}, {Zhang}  \& {Li}}{{Zhang}
  et~al.}{2014}]{zhangIntroductionHardXray2014}
{Zhang} S.,  {Lu} F.~J.,  {Zhang} S.~N.,   {Li} T.~P.,  2014, in {Takahashi}
  T.,  {den Herder} J.-W.~A.,   {Bautz} M.,  eds,  Society of Photo-Optical
  Instrumentation Engineers (SPIE) Conference Series Vol. 9144, Space
  Telescopes and Instrumentation 2014: Ultraviolet to Gamma Ray. p. 914421,
  \mn@doi{10.1117/12.2054144}

\bibitem[\protect\citeauthoryear{{Zhang} et~al.,}{{Zhang}
  et~al.}{2018}]{zhangInsightHXMTMissionIts2018}
{Zhang} S.,  et~al., 2018, in {den Herder} J.-W.~A.,  {Nikzad} S.,   {Nakazawa}
  K.,  eds,  Society of Photo-Optical Instrumentation Engineers (SPIE)
  Conference Series Vol. 10699, Space Telescopes and Instrumentation 2018:
  Ultraviolet to Gamma Ray. p. 106991U (\mn@eprint {arXiv} {1910.04434}),
  \mn@doi{10.1117/12.2311835}

\bibitem[\protect\citeauthoryear{{Zhang} et~al.,}{{Zhang}
  et~al.}{2020a}]{zhangOverviewHardXray2020}
{Zhang} S.-N.,  et~al., 2020a, \mn@doi [Science China Physics, Mechanics, and
  Astronomy] {10.1007/s11433-019-1432-6}, \href
  {https://ui.adsabs.harvard.edu/abs/2020SCPMA..6349502Z} {63, 249502}

\bibitem[\protect\citeauthoryear{{Zhang} et~al.,}{{Zhang}
  et~al.}{2020b}]{zhangComparisonSimulatedBackgrounds2020}
{Zhang} J.,  et~al., 2020b, \mn@doi [\apss] {10.1007/s10509-020-03873-8}, \href
  {https://ui.adsabs.harvard.edu/abs/2020Ap&SS.365..158Z} {365, 158}

\makeatother
\end{thebibliography}








\bsp	
\label{lastpage}
\end{document}